\documentclass[11pt]{article}
\usepackage{authblk}
\usepackage{epsf}
\usepackage{graphicx}
\usepackage{a4}
\usepackage{amsmath}
\usepackage{amssymb}
\usepackage{cite}		
\usepackage{color}
\usepackage{verbatim} 
\usepackage{empheq} 
\usepackage{tikz}

\usepackage{geometry} 

\def\titlepage{\@restonecolfalse\if@twocolumn\@restonecoltrue\onecolumn
     \else \newpage \fi \thispagestyle{empty}\c@page\z@
        \def\thefootnote{\fnsymbol{footnote}}
	\setcounter{page}{0} }
\def\endtitlepage{\if@restonecol\twocolumn \else  \fi
        \def\thefootnote{\arabic{footnote}}
        \setcounter{footnote}{0}}  
        
\definecolor{c1}{rgb}{1, 0, 0}
\definecolor{c2}{rgb}{0, 1, 0}
\definecolor{c3}{rgb}{0, 0, 1}
\definecolor{c4}{rgb}{1, 0, 1}
\definecolor{c5}{rgb}{0, 1, 1}

\def\etc{\hbox{\it etc.}}
\def\ie{\hbox{\it i.e.}}
\def\nn{\nonumber}
\def\beq{\begin{equation}}
\def\eeq{\end{equation}}
\def\bea{\begin{eqnarray}}
\def\eea{\end{eqnarray}}
\def\EQ{\begin{equation}}
\def\EN{\end{equation}}

\allowdisplaybreaks

\begin{document}

\title{Low-temperature universal dynamics of the  bidimensional \\
Potts model in the large $q$ limit 
}

\author{Francesco Chippari$^1$, Leticia F. Cugliandolo$^{1,2}$ and Marco Picco$^1$}
\affil{\small$^1$\textit{Sorbonne Universit\'e, CNRS UMR 7589, Laboratoire de Physique Th\'eorique et Hautes Energies, 
4 Place Jussieu, 75252 Paris Cedex 05, France}}
\affil{\small$^2$\textit{Institut Universitaire de France, 1, rue Descartes, 75231 Paris Cedex 05, France}}

\date{\today}

\maketitle

\abstract{
We study the low temperature quench dynamics of the two-dimensional Potts model in the limit of large number of 
states, $q\gg 1$. We identify a $q$-independent crossover temperature (the pseudo spinodal)  below which no 
high-temperature metastability stops the curvature driven coarsening process. At short length scales, the latter is
decorated by freezing for some lattice geometries, notably the square one. With simple 
analytic arguments we evaluate the relevant time-scale in the coarsening regime, 
which turns out to be of Arrhenius form and independent of $q$ for large $q$. Once
taken into account dynamic scaling is universal.
}

\newpage

\section{Introduction}

The Potts model~\cite{Potts52,Wu82,Baxter82} is one of the best known models of statistical mechanics. It is an extension of the 
Ising model in which the variables are upgraded to take $q$ integer  values.
The model appears in many areas of physics, as well as at its interfaces with other branches of science.
In the physical context, the large $q$ limit of the ferromagnetic model 
is used to describe grain growth, soap froth evolution,  re-crystallization and late stage sitering~\cite{Weaire84,Stavans93,Glazier90} 
(see also~\cite{HH1993} and references therein). 
Mappings  to other celebrated models of statistical mechanics, such as loop and spin ice models~\cite{Baxter76}
opened the way to a myriad of studies in mathematical physics.
The analysis of its critical properties helped developing the conformal field theory  apparatus~\cite{CFT} and, very recently, the bootstrap approach 
has also been applied to this problem~\cite{Rychkov18,Picco19,Saleur20}.
The anti-ferromagnetic Potts model represents the colouring problem of computer science~\cite{Sokal00,Salas01}. 
Other applications in this realm are community detection in complex networks~\cite{Blatt96,Reichardt04,Ronhovde12}
or the inverse problem in biophysics~\cite{Cocco20}. Concomitantly, the Potts model has also been used to mimic a variety of 
 biophyiscal problems, see {\it e.g.}~\cite{Maree07,Garel88}.
The way in which quench randomness affects the order and universality of phase transitions 
was addressed using mostly  the weakly disordered Potts ferromagnets as a paradigm~\cite{Dotsenko95a,Dotsenko95b,Delfino19}.
Mean-field Potts models with strong disorder realize the random first-order phase transition scenario for the glassy arrest~\cite{KiTh88,ThKi88}.
Last but not least, atomic physics realizations of the Potts model have been recently proposed~\cite{Kalinin18} and particle physics 
applications of the same model also appeared in the literature~\cite{bass1999}.

One of the interests of the ferromagnetic Potts model is that beyond a critical value of the number of single spin states, $q_c$, 
the transition is of first order. Thus, the quench dynamics across this phase transition not only involves the more familiar 
coarsening phenomena described by the dynamic scaling hypothesis~\cite{Bray94,Onuki04,Puri09,KrReBe10,HePl10} but 
also the peculiarities of metastability and nucleation~\cite{Gunton83,Binder87,Oxtoby92,Kelton10}. 
It is, therefore, much richer and, quite surprisingly, still far from being fully understood. 

In this paper, and in an accompanying manuscript~\cite{CoCuEsMaPi20}, 
we build upon the analysis of metastability in the large $q$ bidimensional ferromagnetic 
Potts model presented in Ref.~\cite{onofrio}. 
In short, we study the dynamical behaviour after a rapid quench for  sufficiently large $q$  so that the transition is of 
first order. While curvature driven coarsening should be the leading mechanism
for ordering~\cite{Lifshitz62,Ferrero07,Ibanez07b,Petri08,Loureiro10,Loureiro12}, the Potts model dynamics also present low 
temperature freezing~\cite{Sahni,Derrida,Spirin01,Olejarz13,Denholm,Denholm2} and 
metastability close to the critical temperature~\cite{Meunier00,Ferrero09,Ferrero11,BerganzaEPL,Corberi19,onofrio} that may conspire against the 
system reaching equilibrium after the quench. In both works we 
aim to improve our understanding of the interplay between the coarsening process, 
the low temperature freezing  and the metastability close to criticality.
In particular, we study the influence of  the final reduced temperature $T/T_c(q)$
and the number of states $q$ on the dynamic properties. In this work, we identify the crossover temperature below which high-temperature metastability in 
sub-critical quenches is no longer important (the pseudo spinodal~\cite{Ferrero09}) for various lattice geometries, and we
focus on the low temperature freezing, escape from it, and 
scaling in the subsequent coarsening regime. In~\cite{CoCuEsMaPi20}, instead, the 
multi-nucleation mechanism at higher temperatures is investigated.

The layout of the paper is the following. In Sec.~\ref{sec:model} we recall the definition of the Potts model 
and the parameter dependence of the critical temperature on different bidimensional lattices. We also present in this
Section a short summary of our results.
After discussing some general arguments for metastability (Sec.~\ref{sec:metastability}) and freezing (Sec.~\ref{sec:QinfT0})
in the $q\to\infty$ limit, we show the outcome of the  numerical simulations for various $q$ and reduced temperatures $T/T_c(q)$, using
different lattices (all with periodic boundary conditions) in Sec.~\ref{sec:scaling}.
Details on the heat-bath algorithm that we use, and its analysis in the $q\gg 1$ or $T/T_c(q) \ll 1$ limits, 
are given in Sec.~\ref{subsec:heat-bath}. In the last Section we draw  our conclusions.

\section{The model}
\label{sec:model}

The Potts model~\cite{Potts52} is a generalisation of the Ising model in which
the spin variables take $q$ integer values 
(often associated to colours) and are locally coupled ferromagnetically, that is to say, nearest neighbour exchanges
favour equal values of the spins (colours).  The Hamiltonian is 
\begin{equation}
H_J[\{s_i\}] = - \frac{J}{2} \sum_{\langle ij\rangle} \delta_{s_is_j} 
\end{equation}
with $J>0$, $s_i = 1,\dots, q$, and the sum runs over nearest-neighbours on the lattice 
(each bond contributing twice to the sum).
The model undergoes an equilibrium  phase transition at a critical temperature 
that can be of first or second order
depending on $q$ and the dimension of space, $d$. In 
two dimensions, $d=2$,  the transition is of second-order for $2 \leq  q \leq 4$, while  it is of first-order
for $q > 4$. The critical temperature
depends on the coupling strength, $J$, the dimension, $d$, and the coordination of the lattice, $z$. 
On the square lattice, $z=4$ and~\cite{Potts52,Wu82,Baxter82} 
\begin{equation}
T_c^{\rm square} = \frac{J}{\ln(1+\sqrt{q})} \quad\to\quad
T^{\rm square}_c  \simeq \frac{2J}{\ln q} \;\;\;\; \mbox{for} \;\;\;\; q \gg 1
\label{eq:Tc-square}
\end{equation}
($k_B=1$ henceforth).
On the triangular and honeycomb lattices the critical temperatures are
given by implicit expressions~\cite{Wu82}
\begin{eqnarray}
\begin{array}{rll}
\label{eq:tct}
0 \; \; = & x^3 -3 x + 2 -q  & \qquad \mbox{triangular} \;\;\;\;\; z=6
\; , 
\\
\label{eq:tch}
0 \; \; = & x^3 -3 x^2 -3 (q-1) x + 3 q-1 - q^2  & \qquad \mbox{honeycomb} \;\;\; z=3
\; , 
\end{array}
\end{eqnarray}
with $x=e^{\beta_c J}$ and $\beta=1/T$. 
In the large $q$ limit $\beta_c J \gg 1$, implying $x^3  \gg x^2 \gg x$, and 
\begin{eqnarray}
 T^{\rm triang}_c  \simeq  \frac{3J}{\ln q}
 \; , 
 \qquad\qquad
 T^{\rm honey}_c \simeq    \frac{3J}{2\ln q}
 \; . 
\label{eq:Tc-triang-honey}
\end{eqnarray}
Therefore, in the three cases
\begin{equation}
T_c  \simeq  \frac{zJ}{2\ln q} \qquad\quad \mbox{for} \qquad q \gg 1
\label{eq:Tcz}
\end{equation}
and $T_c$ diminishes logarithmically with $q$.

\begin{figure}
\begin{center}
\scalebox{0.7}{%
\begin{tikzpicture}[scale=1.00]
\draw[ thick]  (0.0,0.0) -- (8.0,0.0);
\draw[ thick]  [->] (0.0,0.0) -- (0.0,7.1);
\draw[ thick]  (4.0,0.0) -- (4.0,7.1);
\draw[ thick]  (8.0,0.0) -- (8.0,7.1);
\filldraw (0,0) circle (3pt);
\filldraw (4,0) circle (3pt);
\filldraw (8,0) circle (3pt);
\node at (-0.8,1.5) {$q=10^3$};
%
\node at (-0.8,2.0) {$q=10^4$};
%
\node at (-0.8,2.5) {$q=10^5$};
%
\node at (-0.8,3.) {$q=10^6$};
%
\node at (-0.8,4.5) {$q=10^9$};
%
\node at (-0.8,7.) {$q=\infty$};
\node at (-0.3,-0.3) {$0$};
\node at (4.0,-0.4) {$T=T_c/2$};
\node at (8.3,-0.3) {$T=T_c$};
\fill[color=red!40]
(7.98,1.5) -- (7.98,5.5)
-- (6.8,5.5) -- (6.98,4.5)
-- (7.22,3) -- (7.38,2.5)
-- (7.54,2) -- (7.87,1.5)
-- cycle;
\fill[color=red!40]
(4.02,7.) -- (7.98,7.)
-- (7.98,5.5) -- (6.8,5.5)
-- (6.75,6.) -- (6.575,6.3) 
-- (6.4, 6.5) 
-- (6.1,6.6) -- (5.1, 6.7)
-- (4.5, 6.8) -- (4.2,6.9)
-- cycle;
\fill[color=green!20]
(0.02,1.) -- (3.98,1.)
-- (3.98,7.) -- (0.02,7.) -- cycle;
\filldraw (4.0,7.) circle (3pt);
\node at (2.1,3.5) {Freezing};
\node at (2.1,2.5) { \& Coarsening};
\node at (5.6,4.5) {Metastability, };
\node at (5.6,3.5) { Multinucleation};
\node at (5.6,2.5) { \& Coarsening};
\node at (7.5,5.) {\rotatebox{90}{Metastability}};
\filldraw (0,1.5) circle (1pt);
\filldraw (7.85,1.5) circle (1pt);
\filldraw (0,2.0) circle (1pt);
\filldraw (7.52,2) circle (1pt);
\filldraw (0,2.5) circle (1pt);
\filldraw (7.36,2.5) circle (1pt);
\filldraw (0,3.0) circle (1pt);
\filldraw (7.2,3) circle (1pt);
\filldraw (0,4.5) circle (1pt);
\filldraw (6.96,4.5) circle (1pt);
\end{tikzpicture}
}
\end{center}
\caption{\small A sketch of the phase diagram of the $2d$ Potts square-lattice model. 
The ($T/T_c(q)\leq 1$, $q\gg 4$) plane with the crossover lines between different types of
dynamic behaviour are displayed. The black dots sitting on the limit between the (pink) metastability and (white) multi-nucleation regions 
were obtained in~\cite{onofrio}. This paper focuses on the dynamics in the (light green) regime $T<T_c(q)/2$ while~\cite{CoCuEsMaPi20} concentrates
on the multi-nucleation and further coarsening arising in the white region.
}
\label{fig:sketch1}
\end{figure}
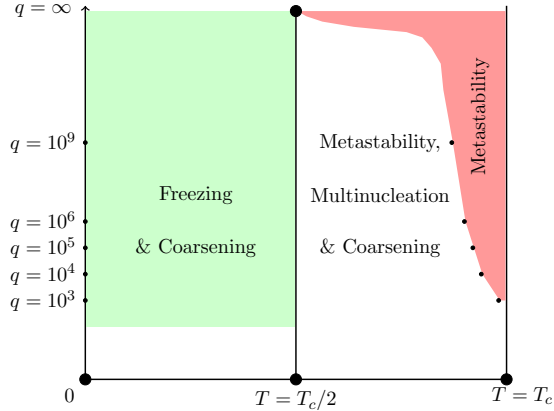

Before entering into the details of our study, let us give here a concise description of our results, found with 
a combination  of analytic arguments and numerical simulations. We use 
systems with $N = L^2$ and $L=10^4$  sites, and  $q$ ranging from $10^2$ to $\infty$.
The various regimes in the $(T,q)$ plane are  sketched  in Fig.~\ref{fig:sketch1}. 
We consider either a subcritical quench starting from a completely disorder configuration or a quench above 
the critical temperature starting from a completely ordered configuration.  
Exploiting the ideas developed in~\cite{onofrio}, we identify a finite temperature interval around the critical one in which the $q\to\infty$ model remains 
metastable (blocked in the initial state forever) after both lower and upper critical quenches. 
The lower limit of this interval is at $T=T_c(q)/2$ with $T_c(q)$ the critical temperature of the corresponding lattice.
For finite $q$ the lifetime of the metastable states, can be extremely long and go beyond any reachable time-span even for not-so-large values of $q$. 
This occurs in the region labeled  ``Metastability''  for the sub-critical quench and painted in pink in Fig.~\ref{fig:sketch1}.
Below the lower limit of this Metastability region, two kinds of mechanisms can lead  to 
equilibration: either multi-nucleation~\cite{CoCuEsMaPi20} followed by coarsening or just curvature driven coarsening. 
We identify the spinodal cross-over 
between the metastable and unstable region at the $q$-independent temperature $T=T_c(q)/2$, the vertical line separating the 
green and white sectors in Fig.~\ref{fig:sketch1}. Finally, at sufficiently low temperatures, 
the systems get quickly trapped in a partially ordered state with life-time 
diverging in the zero temperature and infinite $q$ limits. Only at finite temperature and finite $q$, after leaving these blocked states
at a parameter dependent time-scale that we determine here, 
the dynamics enter the proper curvature driven asymptotic regime. In the latter, the typical length-scale grows algebraically with the 
expected universal power $1/2$ and the prefactor is simply given by a $q$-independent Arrhenius factor.

The main focus of  this paper is the study of the coarsening evolution of the large $q$ Potts model quenched from 
high temperature and, in particular, the analysis of the cross-over from the temporarily blocked states and the coarsening
regime, and its scaling properties. In a companion paper~\cite{CoCuEsMaPi20} we study the metastability and escape from it 
arising closer to the critical temperature, see Fig.~\ref{fig:sketch1}, and the peculiar finite size effects in the 
dynamic evolution.

Concretely, we  measure the time evolution and parameter ($q$ and $T/T_c$) dependence of the growing length, 
$R(t; q, T/T_c)$, which
quantifies the typical linear extent of the ordered patches in the low temperature dynamics
(geometric spin clusters). A way to measure  $R$ is to monitor the energy 
per spin at time $t$, $e(t; q, T/T_c)$, which is associated to the total length of the interfaces.
Accordingly, 
\beq
R(t;q, T/T_c)= \frac{e(t\to\infty; q, T/T_c)}{e(t\to\infty; q, T/T_c)-e(t;q, T/T_c)}
\ .
\eeq
For a random initial condition, $e(t=0)\simeq 0$ for all $q$ and $T/T_c$. 
At very low temperature $e(t\to\infty)\neq 0$ (e.g., for the 
square lattice, $e(t\to\infty) \simeq -2J $) for all $q\geq 2$ and $R(t=0) \simeq 1$. 

The dynamic scaling hypothesis~\cite{Bray94,Onuki04,Puri09,KrReBe10,HePl10}
states that the time-dependence of $R$ should be 
universal in the coarsening regime, and is expected to apply for 
$\delta \ll R \ll L$ with $\delta$ the lattice spacing and $L$ the linear system size. 
This means that if, as in curvature driven coarsening, 
kinetic ordering is ruled by 
the algebraic law $t^{1/2}$, 
the exponent should not depend on the parameters,
and all their dependencies must appear in a pre-factor, in such a way that
\beq
R(t;q, T/T_c) \simeq [\lambda_q(T/T_c) \, t]^{1/2}
\label{eq:hypothesis}
\ .
\eeq
We will examine this guess and confirm that it holds in the asymptotic coarsening 
regime, after leaving any frozen state, in the restricted temperature interval 
\begin{equation}
T/T_c \leq 1/2 \;\; \mbox{in the large} \;\; q \;\;  \mbox{limit}
\; .
\label{eq:limit}
\end{equation}
In the $q\to\infty$ model, at higher though still subcritical temperatures, the system stays blocked in the disordered initial state.
For finite $q$, and $T$ beyond the limit in Eq.~(\ref{eq:limit}), the system performs multi-nucleation before reaching a 
coarsening regime~\cite{CoCuEsMaPi20}. We will also show that for sufficiently 
large $q$ or low reduced temperature $T/T_c$ the parameter
dependence of the time-scale  $\lambda^{-1}_q(T/T_c)$ is 
\beq
t_S(q,T/T_c) = \lambda^{-1}_q(T/T_c) \simeq a \, e^{J/T}  \simeq  a \, q^{2T_c/(zT)}  
\label{eq:time-scale}
\eeq
with $a$ a constant.
Finally, we will prove that, on the square and honeycomb lattices, the growing length~(\ref{eq:hypothesis}) with the time-scale (\ref{eq:time-scale}) establishes 
when the length $R(t;q, T/T_c)$ detaches from a long-lived plateau 
\begin{equation}
R(t;q, T/T_c) \simeq R_p(q,T/T_c) \qquad \mbox{at} \qquad t \simeq t_S(q,T/T_c) \simeq e^{J/T}
\; . 
\end{equation} 
$R_p$ is the typical linear extent of ordered patches in 
the zero temperature blocked states. It is of the order of a few $\delta$ (concretely, $3.63 \, \delta$ for $q\to\infty$ 
on the square lattice) and  it very weakly depends on $q$ and $T/T_c$. The triangular lattice model does not show this kind of
blocking~(see also~\cite{Sahni,Denholm2}).

Therefore,  we conclude that, in quenches to $T/T_c \leq 1/2$, 
after a short transient, the  large $q$ Potts model  reaches a state of the kind of the 
blocked configurations at zero temperature on the square and honeycomb lattices, while it progressively orders with no arrest 
on the triangular lattice. On the first two lattices,  the blocked state survives until a time-scale $t_S$ 
of the order of  $e^{J/T}$ when the dynamics cross over from $R\simeq R_p$ to the conventional curvature driven coarsening, 
$R\simeq R_p \, (t/t_S)^{1/2}$. Consistently, $t_S$ diverges for $T/J\to 0$. 
This study is complemented by the one that 
two of us and collaborators present in~\cite{CoCuEsMaPi20} where we study the dynamics of quenches to $T/T_c>1/2$ for various 
(not that large) values of $q$.

\section{Metastability in the $q\to\infty$ limit}
\label{sec:metastability}

As there exist an ordered and a disordered phase separated by a phase transition at $T_c$ one could naively 
expect that, starting from an initial configuration typical of the ordered phase
and suddenly changing the temperature beyond the critical one, after some (short) transient 
the system should disorder. Conversely, starting from an initial state in the disordered 
phase, and changing the temperature below the critical one, 
the expectation would be that the system orders, at least over long length scales, after some time. However, 
some simple considerations allow us to show that this does not happen everywhere in the high 
temperature phase for the upper critical quench, nor everywhere in the  low temperature phase
for the lower critical one, in the infinite $q$ limit. 

In this limit, the critical temperatures of the square, triangular and honeycomb lattice Potts models
are given by Eq.~(\ref{eq:Tcz}) which can also be expressed as
\begin{equation}
e^{\beta_c J} \simeq q^{2/z}
\label{eq:betacz}
\; . 
\end{equation}

The Boltzmann equilibrium weight of one (out of $q$) fully ordered configurations, one of the ground states, 
is $P_{\rm 1ground} = e^{\beta (zJ/2) N}/Z$, with $Z$ the partition function and $N$ the number of spins. 
A first excited state is obtained from this ground state by changing a single spin to any of the other $q-1$ orientations.
The energy gap is $E_{\rm exc} -E_{\rm 1ground} = zJ$ and the ratio of the two probabilities tends to
$P_{\rm exc}/P_{\rm 1ground} \simeq q e^{-\beta zJ}$ in the large $q$ limit. Accordingly, 
the change of a spin can occur, in the large $q$ limit, 
if and only if $e^{ \beta zJ} < q$. On the square, triangular and honeycomb lattices
we can now use  Eq.~(\ref{eq:betacz}) to prove that the disordering dynamics can be active only for
\begin{equation}
\beta <  \beta_c/2 \;\; \mbox{or} \;\; T > 2 T_c 
\; . 
\end{equation}
Thus, starting from a completely 
ordered state, the system either i) remains blocked in this ground state at any temperature $T < 2 T_c$ or
 ii) disorders completely at $T > 2 T_c$. 

Next, we can consider the case in which the initial state is disordered. Because of the $q\to\infty$ limit, 
typically, each spin takes a different value. The energy  of such a fully disordered configuration vanishes, 
$E_{\rm dis}=0$, and its probability weight is $P_{\rm dis} = q^N/Z$. 
After a quench to a sub-critical temperature, one of the $N$ available spins will try to align with one of its neighbours.
The energy of a state with a ``bond'' is then $E_{\rm bond} = - J$ and its probability $P_{\rm bond} = (e^{\beta J}/q) P_{\rm dis}$.
The condition to start ordering is then 
\begin{equation}
\frac{P_{\rm bond}}{P_{\rm dis}} \geq 1  \quad\Rightarrow \quad 
e^{\beta J}\geq q \simeq e^{\beta_c Jz/2}
\quad \Rightarrow \quad \beta \geq \frac{z\beta_c}{2} \;\;
\mbox{or} \;\; T \leq \frac{2T_c}{z}
\; .
\end{equation}
 
\begin{figure}[!ht]
\vspace{0.25cm}
\begin{center}
\hspace{-2cm} (a) \hspace{4.5cm} (b) \hspace{4cm}
\\
\includegraphics[width=5cm,height=5cm]{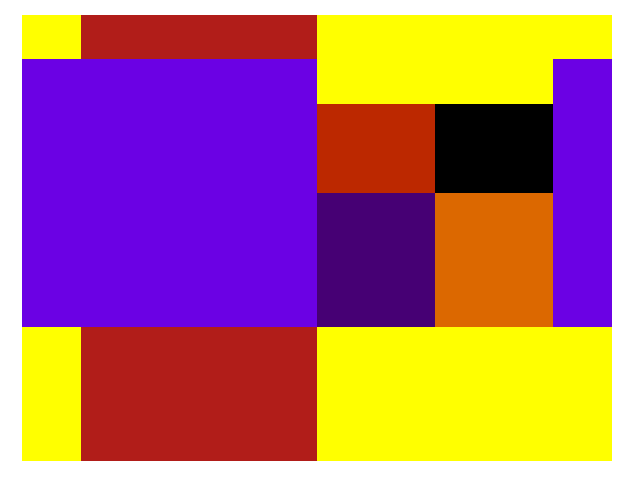}
\includegraphics[width=5cm,height=5cm]{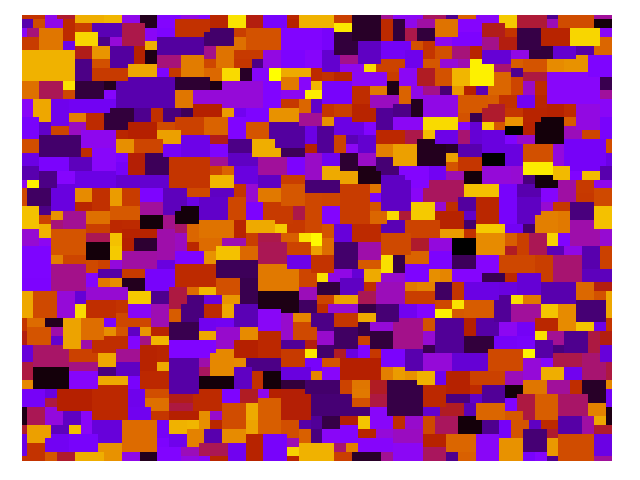}
\caption{\small 
Asymptotic configurations of the Potts model in the infinite $q$ limit after a quench to $T< T_c/2$. 
The lattice is a squared one with  periodic boundary conditions and linear sizes
$L=10$ (a) and $L=10^2$ (b).  The 
dynamic updates follow the heat-bath rules given in Eq.~(\ref{eq:updates-Qinfty}).
}
\label{SQI}
\end{center}
\end{figure}

However, in practice, this does not mean that after a quench to $T \leq 2T_c/z$ the state will order completely. 
For concreteness, let us focus on the square lattice problem.
Indeed, each spin will try to align with one of its neighbours\footnote{We use heat-bath dynamics, more details are given in Sec.~\ref{subsec:heat-bath}.}.
Thus, after a full update of the lattice, all the spins will have created a ``satisfied'' bond with a random neighbour. 
At the next update, any of these bonds can break 
if one of the spins in the pair changes to take the value of another neighbour. 
It is easy to observe that if a spin has the same value as two of its neighbours, $E_{2 {\rm bonds}}=-2J$, it will then be  much more stable than if 
it aligns with only one of its neighbours, $E_{\rm bond}=-J$. If the two bonds form a corner, as shown in the left part of the following sketch, then another spin 
which close a square will  have a large probability to take the same value. Thus, the gray spin will flip to a blue spin, adding two more bonds. 
\vspace{0.25cm}
\begin{center}
\begin{tikzpicture}[scale=0.7]
\node[circle,minimum size=0.3cm,fill=blue] at (0.,1.0) {};
\node[circle,minimum size=0.3cm,fill=blue] at (0.,0.0) {};
\node[circle,minimum size=0.3cm,fill=blue] at (1.,0.0) {};
\node[circle,minimum size=0.3cm,fill=gray] at (1.,1.0) {};
\draw[thick][blue]  (0.,0.) -- (1.,0);
\draw[thick][blue]  (0.,0.) -- (0.,1);
\draw [very thick] [ ->] (2.,0.5) -- (4.,0.5);
\node[circle,minimum size=0.3cm,fill=blue] at (5.,1.0) {};
\node[circle,minimum size=0.3cm,fill=blue] at (5.,0.0) {};
\node[circle,minimum size=0.3cm,fill=blue] at (6.,0.0) {};
\node[circle,minimum size=0.3cm,fill=blue] at (6.,1.0) {};
\draw[thick][blue]  (5.,0.) -- (6.,0);
\draw[thick][blue]  (5.,0.) -- (5.,1);
\draw[thick][blue]  (6.,0.) -- (6.,1);
\draw[thick][blue]  (5.,1.) -- (6.,1);
\end{tikzpicture}
\end{center}
\vspace{0.25cm}
It is then easy to see that squares and rectangles form the more stable small structures. 
Consequently, after a few iterations, the configurations are filled with small squares and rectangles. In particular, one observes 
the existences of so called T-junctions which were identified as the main reason for which blocked states 
occur in finite $q$ Potts models at zero temperature\cite{Lifshitz62,Glazier90,Olejarz13}:
\vspace{0.25cm}
\begin{center}
\begin{tikzpicture}[scale=0.7]
\node[circle,minimum size=0.3cm,fill=blue] at (0.,0.0) {};
\node[circle,minimum size=0.3cm,fill=blue] at (0.,1.0) {};
\node[circle,minimum size=0.3cm,fill=blue] at (1.,0.0) {};
\node[circle,minimum size=0.3cm,fill=blue] at (1.,1.0) {};
\node[circle,minimum size=0.3cm,fill=red] at (2.,0.0) {};
\node[circle,minimum size=0.3cm,fill=red] at (2.,1.0) {};
\node[circle,minimum size=0.3cm,fill=red] at (3.,0.0) {};
\node[circle,minimum size=0.3cm,fill=red] at (3.,1.0) {};
\node[circle,minimum size=0.3cm,fill=green] at (0.,2.0) {};
\node[circle,minimum size=0.3cm,fill=green] at (0.,3.0) {};
\node[circle,minimum size=0.3cm,fill=green] at (1.,2.0) {};
\node[circle,minimum size=0.3cm,fill=green] at (1.,3.0) {};
\node[circle,minimum size=0.3cm,fill=green] at (2.,2.0) {};
\node[circle,minimum size=0.3cm,fill=green] at (2.,3.0) {};
\node[circle,minimum size=0.3cm,fill=green] at (3.,2.0) {};
\node[circle,minimum size=0.3cm,fill=green] at (3.,3.0) {};
\draw[very thick]  (-0.5,1.5) -- (3.5,1.5);
\draw[very thick]  (1.5,-0.5) -- (1.5,1.5);
\end{tikzpicture}
\end{center}
\vspace{0.25cm}
Typical snapshots displaying this 
fact  for $L=10$ and $L=10^2$ are shown in Fig.~\ref{SQI}, at a time $t=10^3$ after a quench of the square lattice 
Potts model towards $T < T_c/2$. In the infinite $q$ limit, the dynamics at $T<T_c/2$ 
are at effectively vanishing temperature, and these configurations are stable. Thus, this run has only partially 
ordered on the square lattice. 

From these simple arguments we conclude that the non-trivial ordering process, 
in the infinite $q$ limit, is restricted to  a quench from a disordered state to $T \leq 2T_c/z$. For
 $2T_c/z < T$ the dynamics are blocked and the
systems remain frozen in the disordered initial state.
Depending on the lattice, the phase ordering kinetics at $T \leq 2T_c/z$ can, however, go through temporarily blocked states 
with interesting patterns, as the ones illustrated in Fig.~\ref{SQI} for the square geometry. These are typical blocked states at
zero temperature, as the ones studied in~\cite{Ibanez07b,Olejarz13,Denholm,Denholm2}. 
After a  $T/T_c$ dependent time-scale needed to escape such states, the system enters the proper dynamic scaling regime that takes it towards 
equilibrium.  

\section{Early approach to temporarily blocked states}
\label{sec:QinfT0}

In this Section, we discuss the similarity of the $q\to\infty$ model quenched to $T/T_c\leq 2/z$
and the finite $q$ model quenched to zero temperature. We first illustrate this feature with some 
numerical data and we then explain the origin of the equivalence by studying in detail the two 
limits of the heat bath transition rates.

\subsection{Numerical method}

We focus on sub-critical quenches. We consider as a starting condition a completely 
disordered configuration,  and we then quench the system to a subcritical  temperature, $T < T_c$, at the initial time $t=0$. 
Next, we start updating with the new temperature. In Monte Carlo simulations, one chooses one site at random 
and changes the value of the spin according to a microscopic stochastic rule. For a system with $N$ spins, $N$ update attempts 
correspond to a single time-step.  We find convenient to use heat-bath dynamics, since each move is  actually
an update in this case. Moreover, a continuous time version~\cite{Bortz75} of the algorithm can be implemented very easily. 
Details on the transition probabilities and their large $q$ and low $T$ limits are given in Sec.~\ref{subsec:heat-bath}.

\subsection{The typical linear size in the blocked states}
\label{subsec:asymptoticRp}

\begin{figure}[!ht]
\vspace{0.5cm}
\begin{center}
\includegraphics[width=8cm]{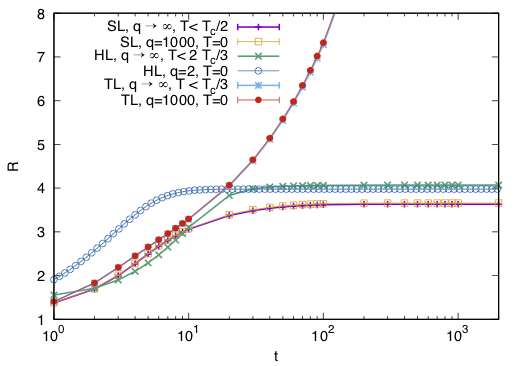}
\end{center}
\caption{\small 
The growing length $R$~vs.~$t$ of the Potts model on the square (SL), honeycomb (HL) and triangular (TL)  lattices with $L=10^3$.
Different curves correspond to values of $q$ and $T$ given in the key. Note the absence of freezing in the triangular lattice case.
}
\label{QI}
\end{figure}

The similarity between the two limits is proven in Fig.~\ref{QI} where we show the growing length $R$ as a function of time
in various cases of interest\footnote{Here and in what follows, we do not write explicitly the $q$ and $T/T_c$
dependence of $R$.}.  In the square lattice model with $q\to\infty$ quenched to $T < T_c/2$
the dynamics eventually block,
in the way illustrated in Fig.~\ref{SQI}, with $R$ approaching, very quickly at $t \simeq 10$, 
a plateau at  $R_p \simeq 3.63$ (we use  
 lattice spacing units such that $\delta=1$).
Besides, the zero temperature quench of a $q=10^3$ model shows a very similar
behaviour, with asymptotically blocked states of the same kind, and roughly the same value of $R_p$. 
We checked that the similarity persists for all $q \geq 10^2$ quenched to $T=0$.

Exceptions to the equivalence are found for sufficiently small $q$. 
 For example, for $q =10$, $R_p$ fluctuates from sample to sample. 
 We did not make a detailed check of all cases between $q=10$ and $q=10^2$, we will simply focus on 
 large enough $q$. Moreover, for large lattices such as $L=10^3$ and $q=10$, even at zero temperature, 
some samples can escape the temporarily blocked configurations and reach a nearly equilibrated state.  

The same equivalence is observed on the honeycomb lattice, with coordination number $z=3$, 
see the other pair of curves in Fig.~\ref{QI}, which approach $R_p \simeq 4$.
The curves show data for the infinite $q$ limit with the condition $T < 2T_c/3$,  and
the  zero temperature quench of the Ising model, $q=2$. 
The latter case was already considered in~\cite{BCPT2017} where it was observed that $R$ saturates 
to a value $R_p \simeq 4$ after a rather short time, $t \simeq 10$. This is due to 
the existence of frozen configurations on the odd-coordinated honeycomb lattice, 
see Sec 6.1 in~\cite{BCPT2017}. Then, for this lattice,
the behaviour at $T < 2T_c/3$ and infinite $q$  is similar to the one at $T=0$ 
for any $q \geq 2$. 

Last, we also show data for the triangular lattice for the the infinite $q$ limit with the condition $T < T_c/3$ and at $T=0$ and $q=10^3$. 
For this lattice, the dynamics do not block. For the largest times,  the growing length $R(t)\simeq t^{1/2}$ 
as expected for standard curvature driven coarsening. The absence of blocking states at zero temperature 
for the triangular lattice is known since a very long time \cite{Sahni,Derrida} and was discussed in a recent work
for small values of~$q$~\cite{Denholm}.

\subsection{Large $q$ or $T\to 0$ limits of the heat-bath rules}
\label{subsec:heat-bath}

For concreteness, we focus on the square lattice case, and we explain, from the behavior of the 
microscopic updates, the origin of the plateau at $R_p \simeq 3.63$
in the growing length curves shown in Fig.~\ref{QI}.

Let us recall the  general rules for the heat-bath dynamics,  on the square lattice, valid for any $q\geq 5$. 
This is done by considering all possible local configurations and their central spin flip evolution.

We follow the scheme introduced in~\cite{onofrio}.
Starting from one spin $S(i)$, we first count  the number of neighbouring spins, $S(j)$,  
taking the same value, $S(j)=S(i)$, and we call this number $n_1$. 
Next, we count the number of neighbours taking other spin values and we organize them in decreasing 
order, $n_2, n_3, \etc$, according to their frequency of appearance. We denote each possible configuration
by $[n_1, n_2, \etc]$ where only the values $n_i \neq 0$ are included. 
On a square lattice there are 12  possible local configurations that we also label with an integer 
$k=0, \dots, 11$
and we write this label between parenthesis.

Now, all possible single spin flip transitions are
\begin{eqnarray}
\begin{array}{llll}
&
\ \, (0) :  [4] \rightarrow  (0), (7)  
& 
\ \,  (1) :  [3,1] \rightarrow (1),  (4),  (8)  
\nn \\
& 
\ \,  (2) : [2,2] \rightarrow  (2),   (2),   (9) 
&
\ \,  (3) : [ 2 , 1 , 1 ]    \rightarrow     (3),  (5),  (10)  
\nn \\
& 
\ \, (4) :  [ 1 , 3 ]  \rightarrow   (4),  (1),  (8)  
&
\ \, (5)  :  [ 1 , 2 , 1 ]    \rightarrow   (5),  (3),   (10)
 \nn  \\
& 
\ \, (6)  :  [ 1 , 1 , 1 , 1 ]  \rightarrow   (6),   (11)  
&
\ \,  (7) :  [ 0 , 4 ]  \rightarrow   (7),   (0)  
\nn \\
& 
\ \, (8) :  [ 0 , 3 , 1 ] \rightarrow   (8),  (1),  (4) 
&
\ \, (9)  :  [ 0 , 2 , 2 ]        \rightarrow     (9),  (2)  
\nn \\
&
(10) :  [ 0 , 2 , 1 , 1 ]  \rightarrow  (10), (3), (5) 
\qquad\qquad
&
(11) :  [ 0 , 1 , 1 , 1 , 1 ] \rightarrow    (11), (6)   
\nn
\end{array}
\end{eqnarray}
and the (0)-(11) states were represented in a figure in Sec. 3.2 in~\cite{onofrio}.

The reasoning behind these formul\ae \ is the following.
The first configuration, named $(0)$, corresponds to one spin surrounded by four neighbors
with the same colour. The central spin can either keep the same value, 
thus the $(0)$ on the right of the arrow, or flip to another value, thus the new configuration $(7) : [0,4]$. 
The local configuration  $(0)$ will remain the same with probability $\simeq e^{4\beta J}$ and change with 
probability $e^{0}=1$ for each other possible value 
of the flipped spin. There are $q-1$ such values. Then normalising the transition probabilities and writing them, for simplicity,
for $J=1$, we have
\beq
P_{0 \rightarrow 0} = \frac{e^{4\beta}}{e^{4\beta}+q-1}  \; , \qquad\qquad
P_{0 \rightarrow 7} = \frac{q-1}{e^{4\beta}+q-1} 
\; . 
\eeq
Following the same kind of analysis, we find that all other transition probabilities are 
\bea
\begin{array}{lll}
P_{1 \rightarrow 1} = \dfrac{e^{3\beta}}{e^{3\beta}+e^{\beta}+q-2}   
& \! \!
P_{1 \rightarrow 4} = \dfrac{e^{\beta}}{e^{3\beta}+e^{\beta}+q-2} 
& \! \!
P_{1 \rightarrow 8} = \dfrac{q-2}{e^{3\beta}+e^{\beta}+q-2}  
\nn\\
P_{2 \rightarrow 2} =  \dfrac{2 e^{2\beta}}{2 e^{2\beta}+q-2}   
& \! \!
P_{2 \rightarrow 9} =  \dfrac{q-2}{2 e^{2\beta}+q-2} 
& \! \!
\nn\\
P_{3 \rightarrow 3} = \dfrac{e^{2\beta}}{e^{2\beta}+2 e^{\beta}+q-3}  
& \! \!
P_{3 \rightarrow 5} = \dfrac{2  e^{\beta}}{e^{2\beta}+2 e^{\beta}+q-3} 
& \! \!
P_{3 \rightarrow 10} = \dfrac{q-3}{e^{2\beta}+2 e^{\beta}+q-3} 
\nn  \\
P_{4 \rightarrow 4} = \dfrac{e^{\beta}}{e^{\beta}+e^{3\beta}+q-2}  
& \! \!
P_{4 \rightarrow 1} = \dfrac{e^{3\beta}}{e^{\beta}+e^{3\beta}+q-2}  
& \! \!
P_{4 \rightarrow 8} = \dfrac{q-2}{e^{\beta}+e^{3\beta}+q-2} 
\nn \\
P_{5 \rightarrow 5} = \dfrac{2 e^{\beta}}{2 e^{\beta}+  e^{2\beta}+q-3}  
& \! \!
P_{5 \rightarrow 3} = \dfrac{e^{2\beta}}{2 e^{\beta}+  e^{2\beta}+q-3}  
& \! \!
P_{5 \rightarrow 10} = \dfrac{q-3}{2 e^{\beta}+  e^{2\beta}+q-3}  
\nn \\
P_{6 \rightarrow 6}  = \dfrac{4 e^{\beta}}{4 e^{\beta}+q-4}  
& \! \!
P_{6 \rightarrow 11} = \dfrac{q-4}{4 e^{\beta}+q-4}  
& \! \!
\nn \\
P_{7 \rightarrow 7} = \dfrac{q-1}{e^{4\beta}+q-1}  
& \! \!
P_{7 \rightarrow 0} = \dfrac{e^{4\beta}}{e^{4\beta}+q-1}
& \! \!
 \nn \\
P_{8 \rightarrow 8} =  \dfrac{q-2}{e^{3\beta}+e^\beta+q-2}  
& \! \!
P_{8 \rightarrow 1} = \dfrac{e^{3\beta}}{e^{3\beta}+e^\beta+q-2} 
& \! \!
P_{8 \rightarrow 4} = \dfrac{e^\beta}{e^{3\beta}+e^\beta+q-2}  
\nn\\
P_{9 \rightarrow 9}  = \dfrac{q-2}{2 e^{2\beta}+q-2}  
& \! \!
P_{9 \rightarrow 2} = \dfrac{2 e^{2\beta}}{2 e^{2\beta}+q-2} 
 \nn \\
P_{10 \rightarrow 10}  = \dfrac{q-3}{e^{2\beta}+ 2 e^\beta + q-3}  
& \! \!
P_{10 \rightarrow 3} = \dfrac{e^{2\beta}}{e^{2\beta}+ 2 e^\beta + q-3}  
& \! \!
P_{10 \rightarrow 5} = \dfrac{2 e^{ \beta}}{e^{2\beta}+ 2 e^\beta + q-3} 
 \nn \\
P_{11 \rightarrow 11}  = \dfrac{q-4}{4 e^{\beta}+q-4}  
& \! \!
P_{11 \rightarrow 6} = \dfrac{4 e^{\beta}}{4 e^{\beta}+q-4} 
& \! \!
\nn
\end{array}
\eea

In the large $q$ limit, the transition probabilities simplify considerably.  
Since $e^\beta= e^{\beta_c T_c/T} =(1+\sqrt{q})^{T_c/T} \simeq 
q^{T_c/2T}$, we deduce that, in this limit, 
$P_{11 \rightarrow 11} $ goes to i) $0$ for $T< T_c/2$, and ii) $1$ for $T_c/2 < T < T_c$.  
Similar simplifications apply to the other probabilities with $(6)$ and $(11)$ states while the other probabilities take values 
$0$ or $1$ for $T<T_c$.  In summary, the limits of the transition probabilities are
\begin{equation}
\begin{aligned}
& P_{0 \rightarrow 0} = 1  \; \qquad\qquad\quad \; P_{0 \rightarrow 7} = 0 
\\
& P_{1 \rightarrow 1} = 1 \; \qquad \qquad\quad  \; P_{1 \rightarrow 4} = 0
\; \qquad\qquad \; P_{1 \rightarrow 8} = 0 
 \\
& P_{2 \rightarrow 2} = 1 \; \qquad\qquad\quad  \;  P_{2 \rightarrow 9} = 0 
\\
& P_{3 \rightarrow 3} = 1  \; \qquad \qquad\quad  \; P_{3 \rightarrow 5} = 0 \; \qquad\qquad \; P_{3 \rightarrow 10} = 0 
\\
& P_{4 \rightarrow 4} = 0  \; \qquad \qquad\quad  \; P_{4 \rightarrow 1} = 1  \; \qquad\qquad \; P_{4 \rightarrow 8} =0  
\\
& P_{5 \rightarrow 5} = 0  \; \qquad \qquad\quad  \; P_{5 \rightarrow 3} = 1 \; \qquad\qquad \; P_{5 \rightarrow 10} = 0  
\\
& P_{6 \rightarrow 6} = 1\ /\ 0  \; \qquad\quad  \;\; P_{6 \rightarrow 11} = 0\ /\ 1 
\\
& P_{7 \rightarrow 7} = 0  \; \qquad \qquad\quad  \; P_{7 \rightarrow 0} = 1  
\\
& P_{8 \rightarrow 8} = 0  \; \qquad \qquad\quad  \; P_{8 \rightarrow 1} = 1 \; \qquad\qquad \;\; P_{8 \rightarrow 4} = 0   
\\
& P_{9 \rightarrow 9} = 0 \; \qquad \qquad\quad  \; P_{9 \rightarrow 2} = 1   
\\
& P_{10 \rightarrow 10} = 0 \; \qquad\quad\quad \; \; P_{10 \rightarrow 3} = 1 \; \qquad\qquad \; P_{10 \rightarrow 5} = 0  
 \\
& P_{11 \rightarrow 11} = 0\ /\ 1  \; \qquad\quad   P_{11 \rightarrow 6} = 1\ /\ 0  
\end{aligned}
\label{eq:updates-Qinfty}
\end{equation}
The two values of the transition probabilities involving $(6)$ and $(11)$ states are written in order for $T< T_c/2$ and $T_c/2 <  T < T_c$. 

Next, we can notice that the rules for $T< T_c/2$ in the large $q$ limit are the same as the ones for finite $q$ 
in the limit $T \rightarrow 0$, thus finding the explanation of the similarity between the two dynamics shown in Sec.~\ref{subsec:asymptoticRp},
see Fig.~\ref{QI}. Let us now give more details on how the evolution takes place in the two cases.

In the large $q$ limit, the initial configuration contains only $(11)$ states. 
If $T> T_c/2$, the state $(11)$ is stable and the system remains disordered forever. If $T <T_c/2$, the $(11)$ states 
change into $(6)$ states. 
In some cases, a spin connecting to another spin is already in a $(6)$ state, this will form a $(3)$ state. 

\vspace{0.25cm}
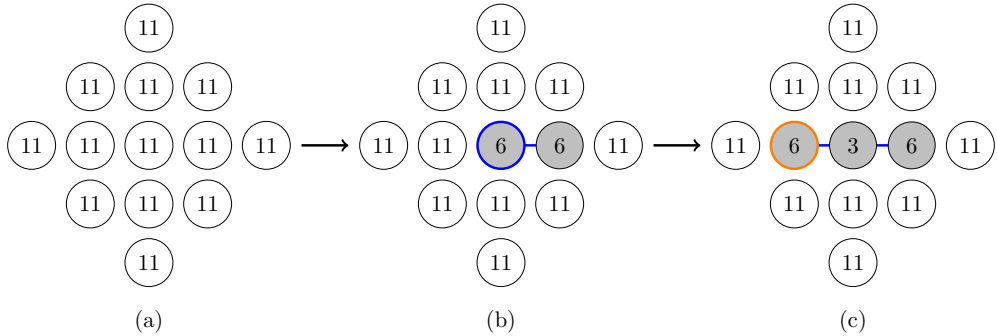
\begin{figure}[!ht]
\begin{center}
\scalebox{0.78}{
	\begin{tikzpicture}
	\node[circle,minimum size=0.8cm,draw=black,fill=white] at (-8.,0.0) {11};
	\node[circle,minimum size=0.8cm,draw=black,fill=white] at (-7.,1.0) {11};
	\node[circle,minimum size=0.8cm,draw=black,fill=white] at (-7.,0.0) {11};
	\node[circle,minimum size=0.8cm,draw=black,fill=white] at (-7.,-1.0) {11};
	\node[circle,minimum size=0.8cm,draw=black,fill=white] at (-6.,2.0) {11};
	\node[circle,minimum size=0.8cm,draw=black,fill=white] at (-6.,1.0) {11};
	\node[circle,minimum size=0.8cm,draw=black,fill=white] at (-6.,0.0) {11};
	\node[circle,minimum size=0.8cm,draw=black,fill=white] at (-6.,-1.0) {11};
	\node[circle,minimum size=0.8cm,draw=black,fill=white] at (-6.,-2.0) {11};
	\node[circle,minimum size=0.8cm,draw=white,fill=white] at (-6.,-3.0) {(a)};
	\node[circle,minimum size=0.8cm,draw=black,fill=white] at (-5.,1.0) {11};
	\node[circle,minimum size=0.8cm,draw=black,fill=white] at (-5.,0.0) {11};
	\node[circle,minimum size=0.8cm,draw=black,fill=white] at (-5.,-1.0) {11};
	\node[circle,minimum size=0.8cm,draw=black,fill=white] at (-4.,0.0) {11};
	
	\draw [very thick] [ ->] (-3.4,0.0) -- (-2.6,0.0);
	\draw[very thick][blue]  (0.,0) -- (1,0);
	\node[circle,minimum size=0.8cm,draw=black,fill=white] at (-2.,0.0) {11};
	\node[circle,minimum size=0.8cm,draw=black,fill=white] at (-1.,1.0) {11};
	\node[circle,minimum size=0.8cm,draw=black,fill=white] at (-1.,0.0) {11};
	\node[circle,minimum size=0.8cm,draw=black,fill=white] at (-1.,-1.0) {11};
	\node[circle,minimum size=0.8cm,draw=black,fill=white] at (0.,2.0) {11};
	\node[circle,minimum size=0.8cm,draw=black,fill=white] at (0.,1.0) {11};
	\node[circle,minimum size=0.8cm,draw=blue,very thick,fill=lightgray] at (0.,0.0) {6};
	\node[circle,minimum size=0.8cm,draw=black,fill=white] at (0.,-1.0) {11};
	\node[circle,minimum size=0.8cm,draw=black,fill=white] at (0.,-2.0) {11};
	\node[circle,minimum size=0.8cm,draw=white,fill=white] at (0.,-3.0) {(b)};
	\node[circle,minimum size=0.8cm,draw=black,fill=white] at (1.,1.0) {11};
	\node[circle,minimum size=0.8cm,draw=black,fill=lightgray] at (1.,0.0) {6};
	\node[circle,minimum size=0.8cm,draw=black,fill=white] at (1.,-1.0) {11};
	\node[circle,minimum size=0.8cm,draw=black,fill=white] at (2.,0.0) {11};
	
	\draw [very thick] [ ->] (2.6,0.0) -- (3.4,0.0);
	\draw[very thick][blue]  (5.,0) -- (6,0);
	\draw[very thick][blue]  (6.,0) -- (7,0);
	\node[circle,minimum size=0.8cm,draw=black,fill=white] at (4.,0.0) {11};
	\node[circle,minimum size=0.8cm,draw=black,fill=white] at (5.,1.0) {11};
	\node[circle,minimum size=0.8cm,draw=orange,very thick,fill=lightgray] at (5.,0.0) {6};
	\node[circle,minimum size=0.8cm,draw=black,fill=white] at (5.,-1.0) {11};
	\node[circle,minimum size=0.8cm,draw=black,fill=white] at (6.,2.0) {11};
	\node[circle,minimum size=0.8cm,draw=black,fill=white] at (6.,1.0) {11};
	\node[circle,minimum size=0.8cm,draw=black,fill=lightgray] at (6.,0.0) {3};
	\node[circle,minimum size=0.8cm,draw=black,fill=white] at (6.,-1.0) {11};
	\node[circle,minimum size=0.8cm,draw=black,fill=white] at (6.,-2.0) {11};
	\node[circle,minimum size=0.8cm,draw=white,fill=white] at (6.,-3.0) {(c)};
	\node[circle,minimum size=0.8cm,draw=black,fill=white] at (7.,1.0) {11};
	\node[circle,minimum size=0.8cm,draw=black,fill=lightgray] at (7.,0.0) {6};
	\node[circle,minimum size=0.8cm,draw=black,fill=white] at (7.,-1.0) {11};
	\node[circle,minimum size=0.8cm,draw=black,fill=white] at (8.,0.0) {11};
	\end{tikzpicture}
	}
\end{center}
\vspace{-0.35cm}
\caption{\small 
	Example of transitions which, starting from a fully disordered configuration (a), 
	partially order the system. In (b), the blue circled spin has changed its value to form 
	a bond with the neighbour on its right. In (c), the orange circled spin has evolved in order 
	to have the same colour as the spin to its right. We can see the appearance of two (6) states in (b) and a (3) state in (c).
	The grey spins have the same colour while the blue bonds indicate a non trivial interaction between spins, as used in \cite{onofrio}. 
}
\label{transition}
\end{figure}

After some iterations, all the states will become $(0), (1), (2)$ or $(3)$ states 
These states can still evolve if a neighbour is flipped but otherwise they are stable. 

For finite $q$ and zero temperature, the dynamics are also very similar. The main difference is that 
the initial configuration contains not only $(11)$ states. 
A simple check shows that the value of the plateau has a weak dependence on $q$. 
Using a lattice with $L=10^3$, we measure $R_p \simeq 3.898 (1)$ for $q=10^2$, 
$R_p = 3.635 (1)$ for $q=10^3$, and 
$R_p = 3.632 (1)$ for $q=10^5$, 
this last value being compatible with the infinite $q$ one. 

A similar analysis can be attempted for the microscopic updating heat-bath rules 
on the honeycomb and triangular cases. For the honeycomb lattice, there exist 
6 states in the heat-bath formulation and such an analysis is even easier than for the square lattice, with 
the 12 states discussed above~\cite{onofrio}. For the triangular lattice, there are 30 states and it
is much more tedious to write a heat-bath  algorithm. Still, in the zero temperature limit, the transition probabilities
simplify and it is manageable. We will show below some results at finite temperature 
using the three lattice geometries. In the triangular lattice case we employed conventional 
Metropolis updates. 

\subsection{Blocked states on the square lattice}
\label{subsec:blocked-states}

\begin{figure}[h!]
\begin{center}
\includegraphics[width=4.6cm,height=4.6cm]{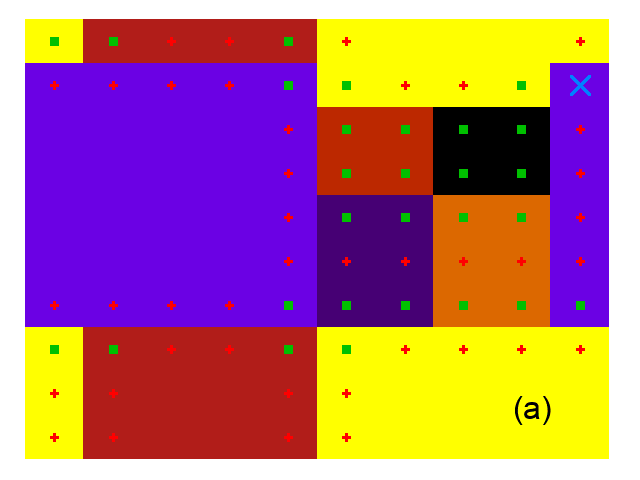}
\hskip -0.3cm
\includegraphics[width=4.6cm,height=4.6cm]{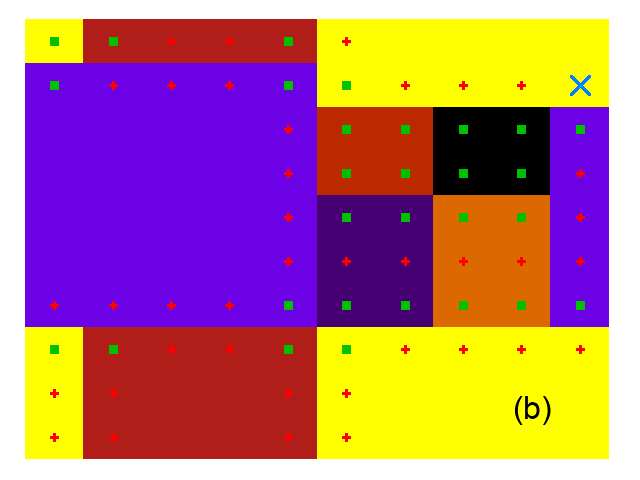}
\hskip -0.3cm
\includegraphics[width=4.6cm,height=4.6cm]{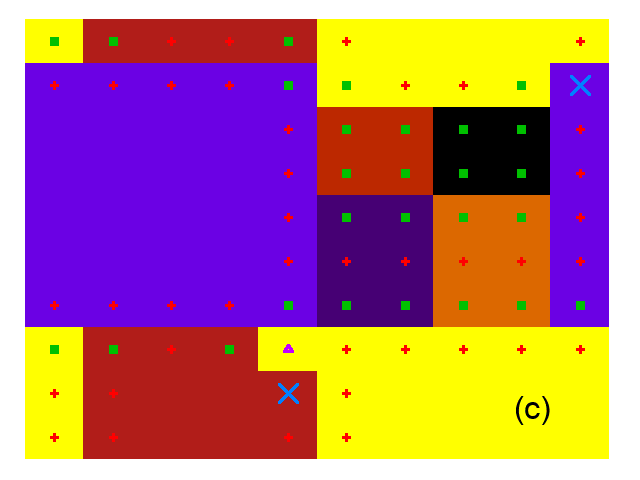}
\includegraphics[width=4.6cm,height=4.6cm]{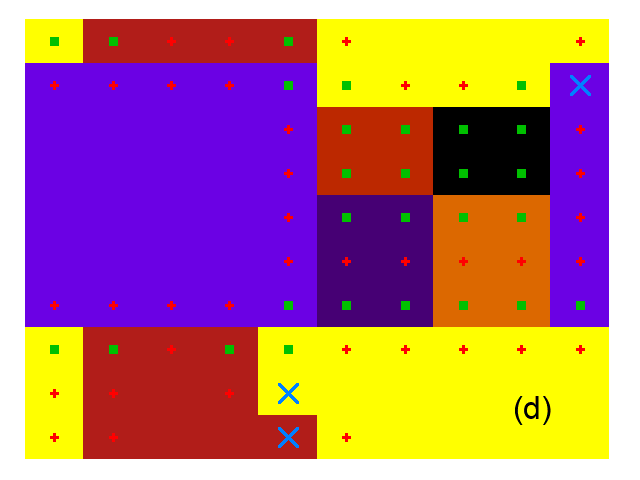}
\hskip -0.3cm
\includegraphics[width=4.6cm,height=4.6cm]{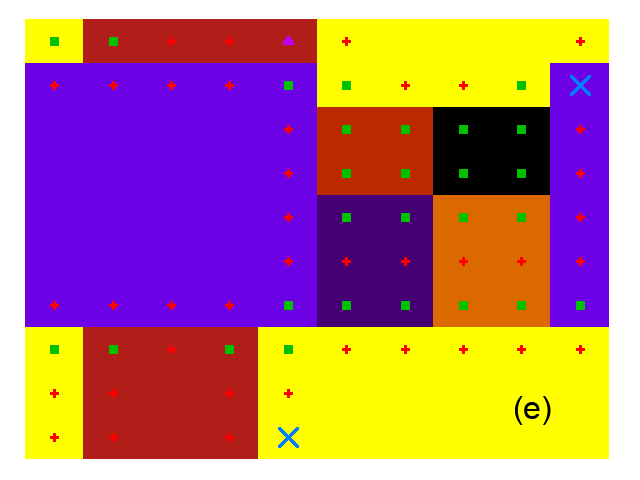}
\hskip -0.3cmx
\includegraphics[width=4.6cm,height=4.6cm]{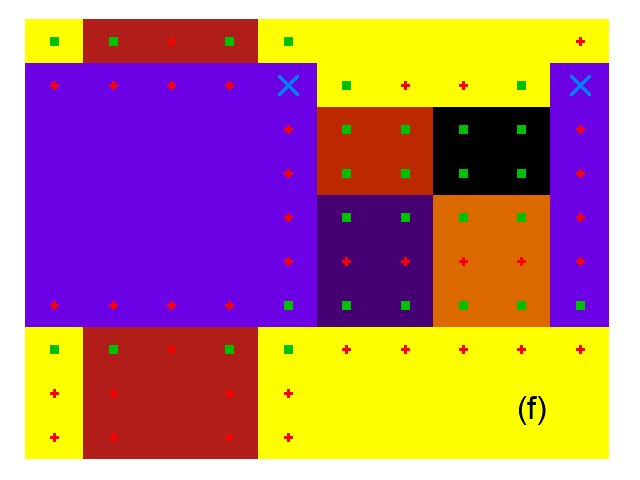}
\caption{\small 
Snapshots of the  Potts model in the large $q$ limit at low temperature. 
The lattice is a squared one with  periodic boundary conditions and
$L=10$. The domains within which the spins take the same value are 
painted with the same color. See the text for details on the convention used to identify the states
of the spins on the domain walls, shown with small data-points with different form and color. The neighboring configurations 
differ by a single spin flip.}
\label{SQI2}
\end{center}
\end{figure}

Consider, again, the blocked configuration obtained after 
a quench from a disordered initial state to $T < T_c/2$  in the $q$ infinite limit of the Potts model on a 
$L=10$ square lattice. 
Such a configuration is shown in the (a) panel of 
Fig.~\ref{SQI2}, where we highlighted the state in which each boundary spin is, using the notation introduced in Sec.~\ref{subsec:heat-bath}. 
For this particular configuration, a direct inspection shows that only four types of states exist, $(0), \ (1), \  (2)$ and $(3)$. 
The $(3)$ states, shown as green squares \textcolor{green}{$\blacksquare$}, 
lie on the corners of the interfaces. They are spins with two neighbours taking the same value and 
two neighbours taking different values from the central one and being also different from each other. 
The $(2)$ states, shown as a blue crosses ({\color{blue}X}), are  blinking states: the central spin has 
two neighbours with the same value and two neighbours with an identical value which is, however, different from the central one. 
For the configuration shown in the (a) 
panel of Fig.~\ref{SQI2}, there is only one blinking state (close to the upper right corner). 
The $(1)$ states, shown as red crosses ({\color{red}+}), are spins at a flat interface with three neighbours being identical to the central one 
and one neighbour taking a different value. The $(0)$ states, corresponding to spins in the bulk of the domains, are not shown. 

Following the transition rules in Eq.~(\ref{eq:updates-Qinfty}), the states $(0), (1)$ and $(3)$ are stable and the corresponding spins cannot 
flip. Only the $(2)$ state can change, producing the configuration shown in the (b) panel of Fig.~\ref{SQI2}. 
Note that under this update, the central spin in the $(2)$ state changes but the state (2) is not modified. 
As a consequence of the flip of the central spin, the four neighbouring spins will also change state but again for stable states:  
two $(1)$ states are changed in $(3)$ states, one $(3)$ state in a $(1)$ state and a $(1)$ state in a $(0)$ state. 
Thus, in this new configuration, only the same spin in the $(2)$ state can change, which will produce again the configuration of the (a) panel in Fig.~\ref{SQI2}. 
In the infinite $q$ limit, the only evolution corresponds to blinking between the configurations shown in the (a) and (b) panels of Fig.~\ref{SQI2}. 

Next, we consider the case of a large but finite $q$ with the condition $e^\beta > q$ corresponding to a low temperature $T < T_c/2$.   
Let us take the configuration in panel (a) of Fig.~\ref{SQI2}. According to the general rules for heat-bath dynamics, a $(0)$ state changes 
to a $(7)$ state with probability $\simeq q e^{-4 \beta}$, 
a $(1)$ state changes to a $(4)$ state with probability $\simeq e^{-2\beta}$ and to a $(8)$ state with  probability $\simeq q e^{-3 \beta}$, 
a $(2)$ state changes to another $(2)$ state with probability $\simeq 1/2$ and to a $(9)$ state with probability $\simeq q e^{-2 \beta}$, 
and a $(3)$ state changes to a $(5)$ state with probability $\simeq e^{-\beta}$ and to a $(10)$ state with probability $\simeq q e^{-2 \beta}$.

In the large $q$ limit with the condition $T < T_c/2$, $e^{-\beta}$ is very small and $qe^{-\beta}<1 $. Then the dominant changes 
are the flips of spins in the $(2)$ state toward another $(2)$ state (thus a blinking) and the next leading changes are the  
spins in the $(3)$ state changed in a $(5)$ state. The flip of a spin in the $(2)$ state will again produce the configuration shown in the (b) 
panel of Fig.~\ref{SQI2}, which will almost surely flip back to the configuration in panel (a), etc. 
Thus this leads to the same blinking behaviour as in the infinite $q$ limit. 
But for a large and finite $q$, after a time $\simeq e^{\beta}$, 
a spin in a $(3)$ state can also flip. Such an example is shown 
in the (c) panel of Fig.~\ref{SQI2}. The spin which changed value (from red to yellow) is now in a $(5)$ state, shown as a purple triangle
(\textcolor{magenta}{$\blacktriangle$}).
We also observe  the appearance of another $(2)$ state just below. Of course, the spin in the $(5)$ state is short lived 
(the probability that it flips back to its previous value is $\simeq 1 - 2e^{-\beta}$). But there exists a finite probability $\simeq 1/2$ 
that the blinking state below is updated first, producing the configuration shown in the (d) panel of Fig.~\ref{SQI2}.
We then see that the $(5)$ state is changed in a (more stable) $(3)$ state, while the blinking state moves down.
Next, there is a finite probability that the new $(2)$ state changes color to produce the configuration seen in the (e) panel of Fig.~\ref{SQI2}. 
This configuration contains again a short lived $(5)$ state. Thus, from this state, there is a large probability 
to end in the configuration shown in the (f) panel of Fig.~\ref{SQI2}.

The conclusion is that starting from configuration (a), one will observe blinking between (a) and (b) configurations until,
after a time $\simeq e^{\beta}$, a transition to a (c) configuration occurs. Next, with a finite probability, the state evolves to 
the (d) configuration and, again with a finite probability, it goes 
to the (e) configuration from which it evolves to the (f) configuration. This shows that the time scale for quitting
the low temperature and large $q$ blocked state is given by 
\begin{equation}
t_S \simeq e^{\beta J} = e^{J/T}
\label{eq:time-scale2}
\end{equation}
where we restored the coupling constant $J$. The same time-scale was found with different arguments by Spirin et al~\cite{Spirin01}
and Ferrero \& Cannas~\cite{Ferrero07}.

Finally, note that this simple scenario is valid for all values of $T$ and $q$ as far as the conditions $e^{-2\beta} \ll e^{-\beta}$ and 
$q e^{-4\beta} \ll e^{-\beta}$ are satisfied. 
This can be rephrased by saying that we expect universal behaviour at 
low temperature and for large $q$. In the next Sections, we will test these predictions with numerical simulations. 
We stress here that a very similar phenomenology is found on the honeycomb lattice and a slightly different one, 
with no blocked states and no plateau in $R$, on the triangular one also to be discussed below.

\section{The growing length}
\label{sec:scaling}

In this Section we study the parameter  ($q$ and $T/T_c$) dependence of the growing length 
$R$ with the aim of proving the hypothesis (\ref{eq:hypothesis}) and finding the explicit form of the pre-factor 
$\lambda_q(T/T_c)$. In most of this Section we work with the square lattice. By the end of it, we 
present data for honeycomb lattice, which confirms the same kind of universality found on the square one. 
We have considered the triangular lattice too but, in this case, we observe a different behaviour.

\subsection{Parameter dependence}

\begin{figure}[!ht]
\begin{center}
\hspace{-3cm} (a) \hspace{6cm} (b)
\\
\includegraphics[width=6.6cm]{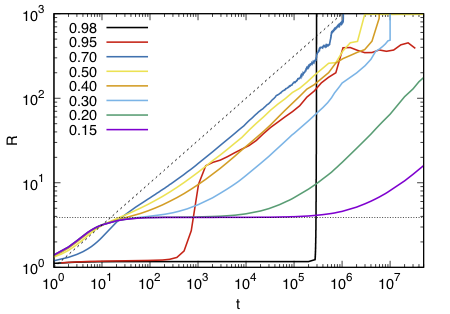}
\includegraphics[width=6.6cm]{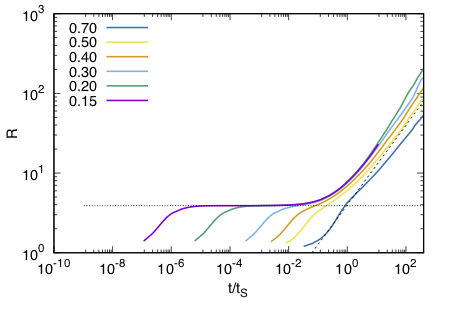}
\\
\hspace{-3cm} (c) \hspace{6cm} (d)
\\
\includegraphics[width=6.6cm]{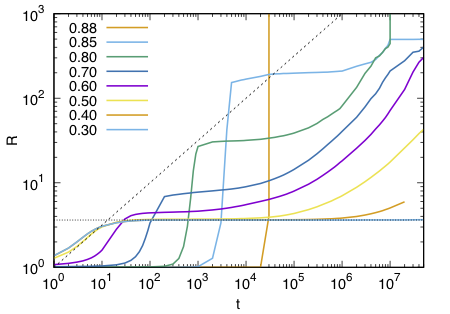}
\includegraphics[width=6.6cm]{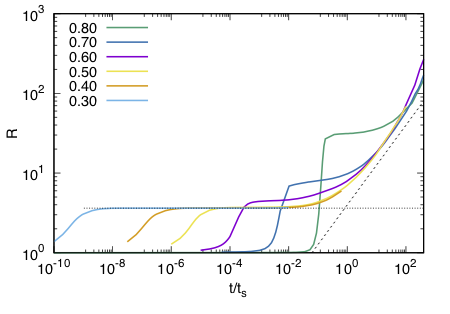}
\end{center}
\caption{\small The growing length 
$R$~vs.~$t$ for $q=10^2$ (a) and (b), 
and $10^6$ (c) and (d) in a square lattice system with  $L=10^3$ and various 
values of $T/T_c$ written in the keys. In the right panels (b) and (d) 
we show a rescaled version in which time is divided by $t_S=e^{J/T}$. 
The horizontal dotted line is at the 
plateau $R_p \simeq 3.63$ and the inclined dashed line is the expected asymptotic 
$t^{1/2}$ law. 
}
\label{R}
\end{figure}

In the left panels of Fig.~\ref{R}, we show $R$ vs. $t$ for $q=10^2$ and $10^6$ 
(from top to bottom) and the reduced temperatures $T/T_c$ given in the keys. In the 
right panels the time is rescaled by $t_S=e^{J/T}$, the time-scale that we identified in Eq.~(\ref{eq:time-scale2}). 
\begin{figure}[!ht]
\begin{center}
\includegraphics[width=7.5cm]{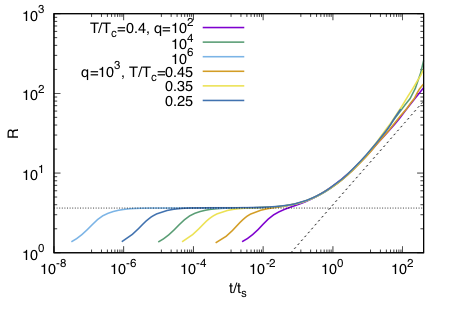}
\end{center}
\caption{\small The growing length $R$~vs.~$t/t_S$ in the square lattice model with $q=10^2, 10^4$ and $10^6$ at $T/T_c=0.40$, 
and with $q=10^3$ at $T/T_c=0.45, 0.35$ and $0.25$. 
The horizontal dotted lines are at the 
plateau $R_p \simeq 3.63$ and the inclined dashed line is the expected asymptotic 
$t^{1/2}$ law. 
}
\label{R2}
\end{figure}

First of all, for both $q$, the curves show the crossover at $T/T_c=1/2$. At very short times, 
the data for $T/T_c>1/2$ demonstrate the early establishment of the (high-temperature) metastable state
with $R\simeq R_m \simeq 1$ and later evolution via the multi-nucleation process~\cite{CoCuEsMaPi20},
while the early evolution of the curves at $T/T_c<1/2$ is temperature independent and rapidly approaches $R_p \simeq 3.7$, 
to only later enter the coarsening regime.

Let us now first discuss the case  $q=10^2$ in more detail. At low relative temperature, up to $T/T_c \simeq 0.30$, 
$R$ makes a small initial jump from $R(0) \simeq 1$ to a finite value $R_p \simeq 3.63$ and keeps this value
 during a time, which we call $t_p$, 
 that increases as we decrease the temperature.  
Next, at later times, there is a crossover towards a regime with the $R(t) \simeq t^{1/2}$ characteristic of the 
standard curvature driven coarsening~\cite{Bray94,Onuki04,Puri09,KrReBe10,HePl10}. 
At higher relative temperatures, there is a similar initial rapid 
increase to $R_p \simeq 3.63$, next an inflection point, and then the coarsening regime. 
This can be seen up to $T/T_c \simeq 0.5$. For even higher relative temperature, 
 $R$ keeps a small value close 
to one, corresponding to the disordered metastable state, and this for longer and longer times $t_m$ 
as we increase temperature. 
For example, for $T/T_c=0.95$, the metastable state survives up to $t_m \simeq 10^2$. 
Next, $R$ increases very rapidly up to a large value $R_l$. 
After this very rapid variation, $R$ increases very slowly first and next faster towards the conventional 
coarsening regime with $R(t) \simeq t^{1/2}$. Note that $R_l$ increases as we increase $T/T_c$, and  
$t_m$ practically diverges at $T/T_c \simeq 0.99$. More details on the 
multi-nucleation processes taking place above $T_c/2$ are given in~\cite{CoCuEsMaPi20}.
 
In summary,  for $q=10^2$ we observe:
\begin{itemize}
\item At $T/T_c \leq 0.5$, there is an initial jump of $R$ to $R_p \simeq 3.63$, a value that 
remains constant up to a time $t_p$ which 
increases as $T/T_c$ decreases. Afterwards, the dynamics reach the coarsening regime with $R$ growing as 
$t^{1/2}$.
\item At $T/T_c > 0.5$, the system remains in the metastable high-temperature state with a very small value $R_m$ up to a time 
$t_m$ which increases with $T/T_c$. At later times, $t>t_m$, we observe a jump towards a finite value $R_l$ 
which increases in a way similar to the increase of $t_m$. 
Next, the typical length grows very slowly, and finally reaches the curvature driven $t^{1/2}$ law. 
\end{itemize}

The situation is similar for other values of $q$, see the other left panel in Figs.~\ref{R}, a system with $q=10^6$. 
The main difference is that for $T/T_c > 0.5$, 
after the jump towards $R_l$, we observe that the slow growth is replaced by a long-lasting plateau as we increase $q$. 
This is particularly clear for large $q$, see the third left panel in Fig.~\ref{R}.

This change of behaviour at $T/T_c \simeq 0.5 = 2/z $ is even better seen if we rescale  time by the time-scale 
$t_S(q,T/T_c) = e^{J/T}$, determined in the previous section. In the  right panels of Fig.~\ref{R},
we show $R$ as a function of $t/t_S(q,T/T_c)$. We observe that, for each $q$, $R$ first goes 
to a plateau at temperatures $T/T_c \leq 0.5$ up to a rescaled time $t/t_S \simeq 10^{-2}$ which 
does not depend on $q$ (confirmed by other values of $q$ not shown). For longer times, the plateau will be escaped  
in a universal way and for longer (rescaled) times, 
the coarsening regime will be reached with $R(t) \simeq t^{1/2}$. 
Note that for $q=10^4$ and $q=10^6$, the curves are identical for $T/T_c \leq 0.5$. For $q=10^2$, the scaling is not as good for $T/T_c$ close 
to $0.5$. This is in agreement with the previous observation that the zero temperature behaviour becomes universal for large $q$ and 
with deviations up to $q \simeq 10^2$. 

Then, for $T/T_c < 0.5$, we claim that 
the behaviour of $R$ is universal if we introduce a rescaling of time by $e^{J/T}$
such that\footnote{In~\cite{Loureiro10,Loureiro12} a rather weak dependence of $\lambda_q(T=T_c/2)$ on $q$ was claimed. Differently 
from here, in those papers only small values of $q$, $q=2,3,8$, and the special temperature $T=T_c/2$ were considered.}
\begin{eqnarray}
\label{scale}
R(t;T/T_c, q) \simeq f(e^{-J/T} t) 
\quad
\mbox{with} 
\quad
f(x)  \;\; \simeq \;\; 
\left\{ 
\begin{array}{ll}
3.63 & \;\; \mbox{for} \;\; x \ll 1 \; , 
\\
x^{1/2} & \;\; \mbox{for} \;\; x \gg 1 \; . 
\end{array}
\right.
\end{eqnarray}
As usual with scaling laws, this behavior is restricted to  $R\gg \delta$, with $\delta$ a length scale of the order of the lattice spacing,
and $R \ll L$ with $L$ the system size where equilibration of at least some samples comes into play.
Our main finding is that for large $q$ ($q \geq 10^3$)  and small $T$ ($T/T_c \leq 0.5$), 
after a short transient the system reaches a state equivalent to the blocked state 
at zero temperature, with $R_p$ determined by the lattice geometry and microscopic dynamics,
and that this blocked state survives up to $t_p \simeq R_p^2 \, e^{J/T} = R_p^2 \, t_S$  when 
the dynamics crosses over to the conventional coarsening one. 
The behavior on the triangular lattice is different and we discuss it below.
A much more detailed analysis of the nucleation process and further phase ordering kinetics in the 
cases $T/T_c(q)>1/2$ is reserved to Ref.~\cite{CoCuEsMaPi20}.

\subsection{Snapshots}

\begin{figure}
\begin{center}
\includegraphics[width=13cm,height=10.5cm]{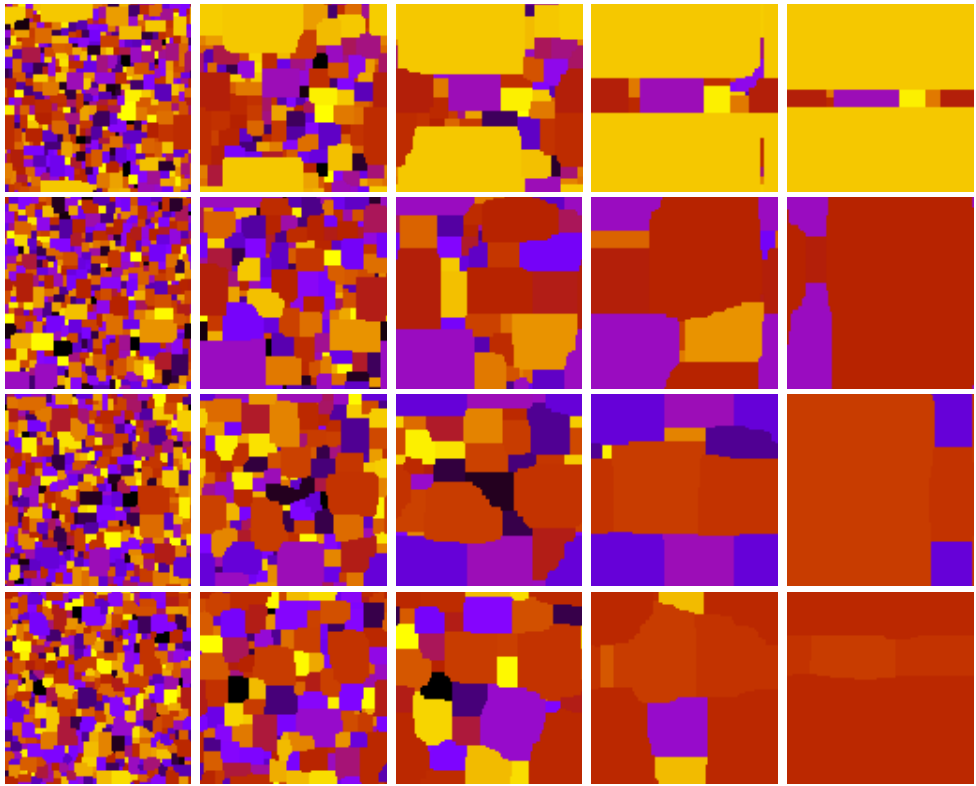}
\end{center}
\caption{\small 
Snapshots at $R(t)=5, 10, 20, 40$ and  $80$ (left to right), $L=10^2$ and $q=10^2$.  From top to bottom $T/T_c = 0.2,  0.3,
0.4$ and $0.5$.
}
\label{FT1}
\end{figure}

In the previous subsection, we argued that there is universal behaviour as a function of temperature after a proper rescaling.
This was shown by considering the behaviour of the growing length $R(t;q,T/T_c)$. We want to confirm this result by 
showing some snapshots of a system with $L=10^2$ and $q=10^2$, see Fig.~\ref{FT1}. 
We present  the instantaneous configurations at the times at which $R=5, 10, 20, 40$ and $80$ 
from left to right, as reached at the relative temperatures 
$T/T_c=0.2, 0.3, 0.4, 0.5$ from top to bottom. The main observation is that the snapshots 
look very much the same for fixed value of $R(t)$. We have also checked similar snapshots for other values of the 
 number of colours, $q=10^3$ up to $10^6$, and 
they also look the same for the same $R$ and relative temperatures $T/T_c$. 

\subsection{Honeycomb lattice}
\label{subsec:honeycomb}

\begin{figure}[!ht]
\vspace{0.25cm}
\begin{center}
$\;$
(a) \hspace{6.5cm} (b) \hspace{3cm} $\;$
\\
\includegraphics[width=6.6cm]{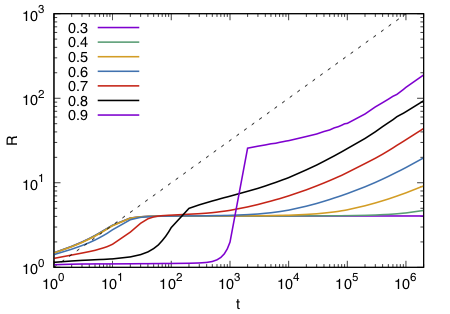}
\includegraphics[width=6.6cm]{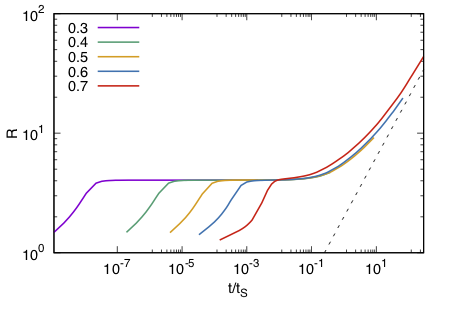}
\end{center}
\vspace{-0.25cm}
\caption{\small The time and  temperature dependence of the growing length
$R$ in the Potts model with $q=10^4$ on the
honeycomb lattice with periodic boundary conditions and $L=10^3$. 
In (a) bare data for many $T/T_c(q)$ given in the key. In (b) 
$R$ against $t$ rescaled by  $t_S = e^{J/T}$  for four temperatures $T<2T_c/z$ and also one temperature   $T>2T_c/z$ which 
approaches the asymptotic $t^{1/2}$ without scaling at short times.  
The dashed inclined line is the curvature driven law $t^{1/2}$.
}
\label{RH}
\end{figure}

To check the universality of our results  over the lattice kind we now show some measurements using 
a honeycomb lattice\footnote{We build the honeycomb lattice starting 
from a square lattice of linear size $L\times L$. At each site, the spin is connected to the left and right, and alternating, 
to the upper or lower raw. For a more detailed description, see \cite{BCPT2017}.}. 
We recall that the critical temperature is given by $e^{\beta_c J} \simeq q^{2/3}$ in the large $q$ limit.
In Fig.~\ref{RH}, we study the time and reduced temperature dependence of the growing length $R$
in a system with $L=10^3$ and $q=10^4$. In this case 
$T_c \simeq 0.162$. The time-dependence of $R$ is similar 
to the one found on the square lattice with the same $q$, see Fig.~\ref{R}. 
Again, at low temperatures, \ie\ $T < 2 T_c/3$, $R$ first 
goes to a plateau at $R_p \simeq 4$, the same value obtained in the infinite $q$ limit after 
a quench towards $T < 2 T_c/3$, see Fig.~\ref{QI}. 
In the right part of Fig.~\ref{RH}, one can see that this plateau exists during a time $\simeq 
e^{J/T}$ (as was the case on the square lattice for $T < T_c/2$). At reduced temperatures above $2T_c/z$
 the system remains in the high temperature metastable state with  $R\simeq 1$  until a sudden jump
 in $R$ towards $R_l$ takes it out of it. $R_l$ increases with $T$.

\subsection{Triangular lattice}
\label{subsec:triangular}

\begin{figure}[!ht]
\begin{center}
\includegraphics[width=7cm]{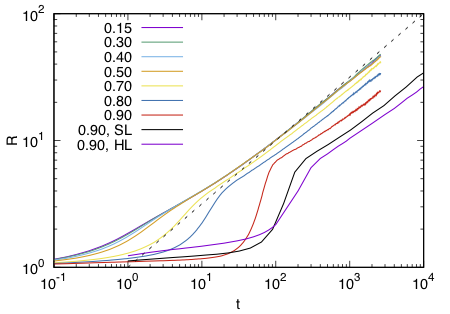}
\end{center}
\vspace{-0.25cm}
\caption{\small 
The growing length  $R$~vs.~$t$ for the triangular  lattice model with $q=10^2$ and $L=10^3$ at different values of $T/T_c$ given in the key.
For comparison, we also show results for the square lattice (SL) and the honeycomb lattice (HL) at $T/T_c=0.90$. The dashed inclined
line is the curvature driven law $t^{1/2}$.
}
\label{Compare}
\end{figure}

We now consider the case of the triangular lattice with coordination number $z=6$. 
In the large $q$ limit,
 $e^{\beta_c J} \simeq q^{1/3}$ and the interesting regime is the one of  quenches below $T_c/3$. 
 Differently from what observed on the square and honeycomb lattices, in such quenches there is 
 no plateau in $R$ and the growing length is not strongly slowed down. This was already observed in~\cite{HH1993} 
 for the $q=10^2$ model at $T=0.1$, while $T_c = 0.635$ for this $q$. We show our results, also for $q=10^2$, 
 in Fig.~\ref{Compare} for various values of  $T/T_c$. We observe that $R$ does not depend on $T$ for 
small $T/T_c$ and we do not observe a plateau. For the smallest value $T/T_c=0.15$, $R$ 
takes the same value as for the zero temperature 
shown in Fig.~\ref{QI} (either at $q=10^3$ or in the infinite $q$ limit). 
Only for $T/T_c =0.4 > T/T_c = 1/3$ we do see a small deviation. 
For large values of $T/T_c$ the behaviour is similar to the one of the other lattices. We have also included, 
in Fig.~\ref{Compare},  the time dependence of $R$  at $T/T_c=0.90$ for the square and 
honeycomb lattice models. 

Note that the measurements using the triangular lattice at a finite temperature have been done 
with a Metropolis algorithm since the heat bath 
one is difficult to implement. We then had to rescale the time by a factor $\simeq q$ to compare the two. 
We found that with a factor $50$, the behaviour of $R$ is similar for the three lattices at $T/T_c=0.90$. 

\section{Conclusions}

Exploiting the $q\to\infty$ limit of the heat-bath Monte Carlo algorithm~\cite{onofrio} we identified 
the temperature interval $[2T_c/z, 2T_c]$ in which high or low temperature initial conditions are metastable after 
sudden sub-critical or upper-critical quenches, respectively, with $z$ the coordination of the lattice. 
In other words, we located the spinodals. 
In the $q\to\infty$ limit, the metastability is ever lasting while,
for finite $q$, the initial states will eventually die out. Once this done, we focused on sub-critical quenches 
for temperatures below the lowest temperatures at which nucleation is observed, that is 
$T<2T_c/z$. For these processes, we showed that on the square and 
honeycomb lattices, after a rapid evolution, the systems temporarily block in configurations that are typical of the asymptotic 
states of the zero temperature dynamics~\cite{Olejarz13,Denholm}. At non-vanishing temperature these states are not fully blocking and the 
systems escape them in a time-scale $t_S \simeq e^{J/T}$ independently of $q$ for large $q$ (see also~\cite{Spirin01,Ferrero07}). 
The proper curvature driven coarsening then takes over 
with the universal  algebraic growing length $R\simeq (t/t_S)^{1/2}$. On the triangular lattice we see no freezing, similarly to what was 
found in~\cite{Sahni,Derrida,Denholm2}, and $R$ is independent of temperature, for $T/T_c < 2/z=1/3$, 
within our numerical accuracy.

In a companion paper~\cite{CoCuEsMaPi20}, the multi-nucleation process in the square lattice model at $T>2T_c/z = T_c/2$ 
was thoroughly studied. The analysis proves that the finite size of the system plays a determinant role in deciding how many 
phases nucleate, with a $\ln L$ growth of this number with the linear system's size. The ordered phases nucleate in
a background that is strongly reminiscent of, and even quantitatively similar to,  the critical state.

\vspace{0.25cm}

\noindent
{\bf Acknowledgements}
We thank F. Corberi, M. Esposito and O. Mazzarisi, with whom two of us performed a parallel study 
of quenches towards parameters close to $T_c(q)$ for which the system shows metastability and 
nucleation~\cite{CoCuEsMaPi20}.

\end{document}